# Decoupling Thermal Properties in Multilayer Systems for Advanced Thermoreflectance Techniques

Tao Chen,[1] and Puqing Jiang [1,*]

[1]*School of Energy and Power Engineering, Huazhong University of Science and Technology, Wuhan, Hubei 430074, China*

**ABSTRACT**. Thermoreflectance techniques, including time-domain thermoreflectance (TDTR), frequency-domain thermoreflectance (FDTR), and the square-pulsed source (SPS) method, are powerful tools for characterizing the thermal properties of bulk and thin-film materials. However, accurately interpreting their signals remains challenging due to intricate interdependencies among experimental variables. This study introduces a systematic framework based on singular value decomposition (SVD) to decouple these interdependent parameters and enhance the reliability of thermal property extraction. By applying SVD to the sensitivity matrix, we identify key parameter combinations and establish essential dimensionless numbers that govern thermoreflectance signals. The framework is applied to a GaN/Si heterostructure, where the performance of TDTR, FDTR, and SPS is evaluated and compared. The results demonstrate a high degree of consistency across all three techniques. Notably, with the intricate relationships of parameters unraveled, TDTR, FDTR, and SPS demonstrate significant potential to simultaneously and accurately extract five to seven key thermal properties, including thermal conductivity, heat capacity, and interfacial thermal conductance of the GaN/Si multilayer system. This framework not only improves the precision of thermoreflectance measurements but also lays a foundation for advanced thermal metrology in research and industrial applications.

## I. INTRODUCTION.

Accurate thermal characterization of materials, particularly at the micro- and nanoscale, is critical for optimizing the performance and reliability of modern electronic, photonic, and energy systems [1-3]. Key thermal properties, such as thermal conductivity, heat capacity, and interfacial thermal conductance, directly impact device functionality in these systems. Thermoreflectance techniques, including time-domain thermoreflectance (TDTR) [4-6] and frequency-domain thermoreflectance (FDTR) [7-9], are widely used for probing these properties in bulk and thin-film materials due to their high spatial and temporal resolution. TDTR, for instance, employs ultrafast laser pulses to generate and monitor thermal responses in samples, enabling the measurement of through-plane thermal conductivity, interfacial conductance, and heat capacity. Meanwhile, FDTR modulates the pump laser at various frequencies, providing a versatile means to assess thermal properties at different depths.

Despite their widespread application, the interpretation of thermoreflectance data remains challenging due to the complex relationships among various experimental variables. In multilayer systems, parameters such as layer thickness, in-plane and cross-plane thermal conductivities, heat capacities, and interfacial thermal resistances are often interlinked, complicating the isolation of individual parameters. Addressing these dependencies is essential to determine the maximum number of reliably extractable variables and to fully leverage the potential of thermoreflectance experiments.

To address this, we employ a framework that applies singular value decomposition (SVD) to thermoreflectance data [10], allowing for a systematic analysis of sensitivity relationships across experimental variables. By examining the sensitivity matrix, this approach identifies key parameter combinations that influence signal behavior, supporting parameter decoupling and enhancing measurement reliability.

To demonstrate the practical utility of this framework, we apply it to a GaN/Si heterostructure, a material system widely relevant for electronic and optoelectronic applications. We compare the parameter extraction capabilities of three thermoreflectance techniques: TDTR, FDTR, and the square-pulsed source (SPS) method, a recently developed technique that combines the strengths of both TDTR and FDTR. The results demonstrate strong consistency across all three techniques, validating the findings of this study. With the intricate relationships clarified, all three techniques exhibit significant potential to simultaneously and accurately extract more parameters.

This study not only advances our understanding of heat transfer mechanisms in multilayer systems but also provides a robust framework for improving the accuracy of thermoreflectance measurements. The methodologies and insights presented have broad implications for material characterization, offering enhanced tools for industries and researchers working on next-generation electronic and thermal management technologies.

* Contact author: jpq2021@hust.edu.cn

## II. BASICS OF THERMOREFLECTANCE EXPERIMENTS

Thermoreflectance methods, including TDTR and FDTR, are widely used to measure thermal properties in both bulk and thin-film materials, especially at micro- and nanoscale dimensions. These techniques rely on the temperature-dependent change in optical reflectance to probe key thermal properties, such as thermal conductivity, heat capacity, and interfacial thermal conductance. Each method offers unique advantages and presents distinct challenges, making them suitable for varying types of thermal analysis.

TDTR is a powerful tool for measuring thermal properties with high temporal resolution. It employs femtosecond laser pulses to generate and monitor thermal responses in materials. [11] Specifically, a pump pulse heats the sample, while a delayed probe pulse measures the corresponding change in reflectance as a function of time delay. This configuration enables precise measurements of through-plane thermal conductivity, interfacial thermal conductance, and heat capacity.

One of the key strengths of TDTR lies in its ability to resolve ultrafast thermal processes and measure thin-film thicknesses with picosecond precision. However, TDTR is limited by its modulation frequency range (typically between 0.1 to 20 MHz), restricting its ability to assess materials with in-plane thermal conductivities below 6 W/(m·K) [12]. Additionally, TDTR systems require precise control over components like the electro-optic modulator (EOM) and mechanical delay stage, making the technique complex and costly. Despite these challenges, TDTR remains a valuable method for high-precision measurements, particularly in layered materials.

FDTR, in contrast, modulates the pump laser at various frequencies and analyzes the phase and amplitude of the resulting thermoreflectance signals. [7] This method eliminates the need for a mechanical delay stage, simplifying the setup and improving stability. By varying the modulation frequency, FDTR can probe thermal properties at different depths, offering a flexible solution for assessing thermal conductivity, interfacial thermal resistance, and heat capacity across a wide range of materials.

FDTR can operate over a broad frequency range, from 1 Hz to 75 MHz, theoretically allowing it to measure materials with a wide range of thermal conductivities. [13] However, FDTR is more susceptible to noise at very low or high frequencies, which can degrade the signal-to-noise ratio and impact measurement quality. In practice, FDTR's performance depends on accurate phase corrections, and environmental factors like electronic noise can significantly affect the quality of the data. Despite these limitations, FDTR's versatility makes it a powerful tool for probing thermal transport phenomena at various scales.

The SPS method, a relatively recent development, combines the key features of both TDTR and FDTR. SPS uses a square-wave modulated pump laser to heat the sample and then record the amplitude of the temperature response over time. [14,15] This method provides time-resolved measurements similar to TDTR while also allowing for a wide range of modulation frequency, from 1 Hz to 10 MHz, akin to FDTR, making it especially effective for measuring low in-plane thermal conductivities, down to about 0.2 W/(m·K).

One of the main advantages of SPS is that it measures the amplitude of the temperature response, which is inherently more stable than the phase signals used in TDTR and FDTR. This resilience to interference enables more reliable extraction of thermal properties, even under challenging experimental conditions. SPS also provides clearer relationships between sensitivity coefficients and experimental parameters, making it well-suited for decoupling complex interactions in multilayer systems.

All three thermoreflectance techniques rely on solving inverse problems to derive thermal properties by fitting experimental data to a thermal diffusion model. Sensitivity analysis plays a crucial role in identifying which parameters can be reliably extracted and in estimating measurement uncertainties. The sensitivity coefficient $S_\xi$, which describes how a small change in a parameter $\xi$ affects the measured signal, guides the fitting process, and provides insights into the coupling between different thermal properties.

In TDTR, FDTR, and SPS, the sensitivity coefficient at each independent variable $x_i$ quantifies how responsive the signal $R$ is to a given parameter $\xi_j$ and is defined as:

$$\left(S_{\xi_j}\right)_{x_i} = \left(\frac{\partial \ln R}{\partial \ln \xi_j}\right)_{x_i} = \left(\frac{\xi_j}{R}\frac{\partial R}{\partial \xi_j}\right)_{x_i} \tag{1}$$

Here, $(S_{\xi_j})_{x_i}$ indicates that a 1% change in $\xi_j$ results in a $(S_{\xi_j})_{x_i}$% change in the signal $R$.

Specifically, in TDTR, the signal $R$ is the negative ratio of in-phase ($V_{\mathrm{in}}$) to out-of-phase ($V_{\mathrm{out}}$) signals, and $x_i$ represents the delay time $t_d$; in FDTR, $R$ can be either the absolute phase $|\phi|$ or the normalized amplitude $A_{\mathrm{norm}}$, and $x_i$ represents the modulation frequency $f$; while in SPS, $R$ is the normalized amplitude $A_{\mathrm{norm}}$, and $x_i$ represents the normalized time $t_{\mathrm{norm}}$.

The logarithmic definition of sensitivity in Eq. (1) is particularly well-suited for signals that vary across several orders of magnitude. Consequently, such signals are typically plotted on a logarithmic scale to enhance visualization.

In FDTR, an alternative linear sensitivity definition is frequently used for the phase signal $\phi$ [7]:

$$\left(S_{\xi_j}\right)_{f_i} = \left(\frac{\partial\phi}{\partial \ln \xi_j}\right)_{f_i} = \left(\xi_j \frac{\partial\phi}{\partial\xi_j}\right)_{f_i} \qquad (2)$$

In this linear definition, $(S_{\xi_j})_{f_i}$ implies that a 1% change in $\xi_j$ results in a change of $(S_{\xi_j})_{f_i} \times 0.01°$ in the phase signal $\phi$. This approach emphasizes signals with large magnitudes, which are generally presented on a linear scale for better clarity.

Figure 1 presents the measurement of a 100 nm Al/sapphire sample using TDTR, FDTR, and SPS, alongside the corresponding sensitivity coefficients for comparison. This thermal system involves 9 parameters, including thermal conductivities ($k_{r,\mathrm{m}}$, $k_{z,\mathrm{m}}$, $k_{r,\mathrm{sub}}$, $k_{z,\mathrm{sub}}$), heat capacities ($C_{\mathrm{m}}$, $C_{\mathrm{sub}}$), thickness ($h_{\mathrm{m}}$), interfacial thermal conductance ($G$), and the laser spot size ($r_0$). Here, $k_r$ and $k_z$ refer to the in-plane and through-plane thermal conductivity, respectively; $C$ is the volumetric heat capacity; and $h$ is the thickness. The subscripts 'm' and 'sub' indicate the properties of the metal transducer layer and the substrate, respectively.

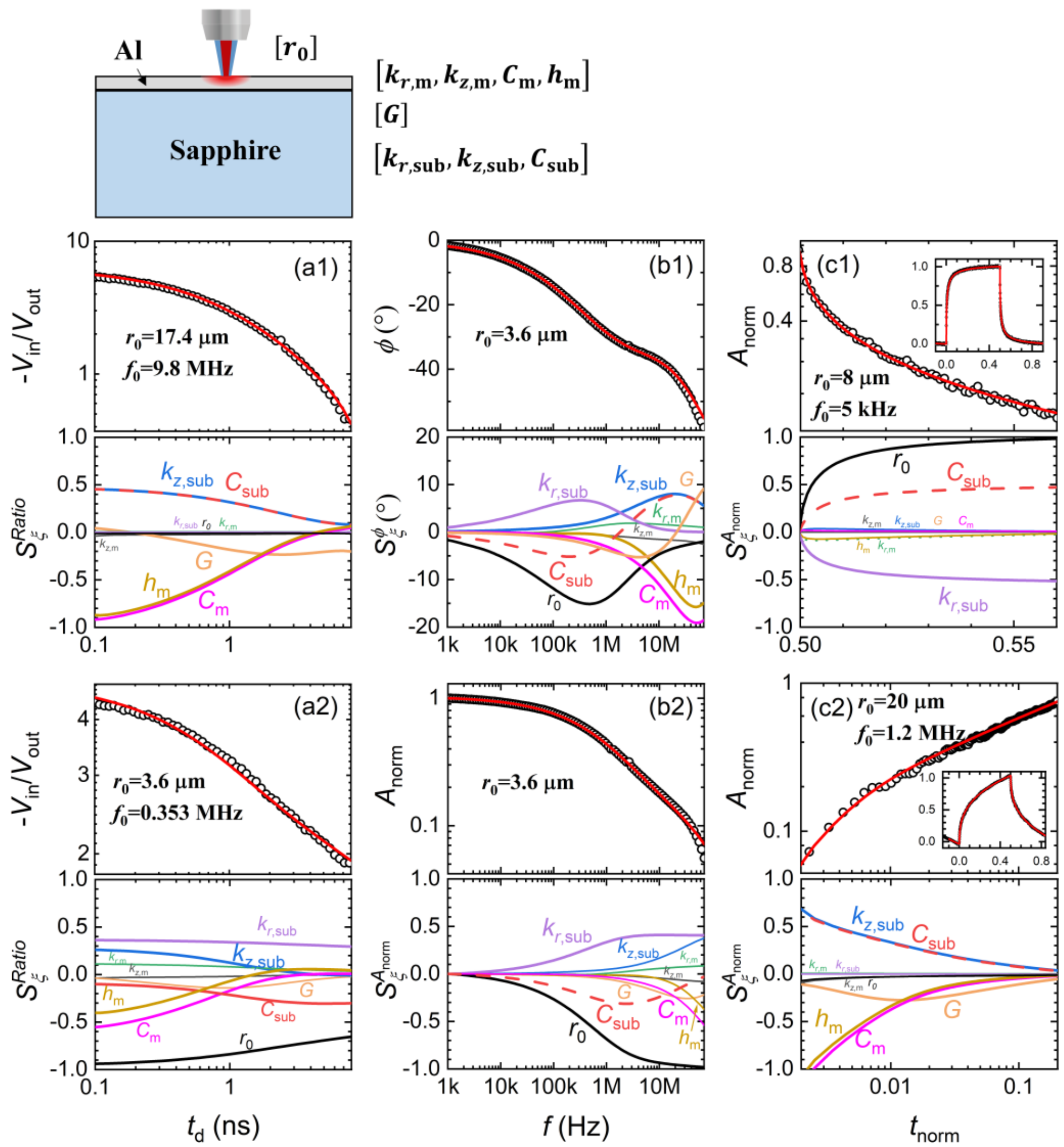

FIG 1. Comparative analysis of thermal measurements on a 100 nm Al/sapphire sample using TDTR, FDTR, and SPS methods. (a1, a2) TDTR signals at 9.8 MHz with a 17.4 μm spot size and 0.353 MHz with a 3.6 μm spot size. Corresponding sensitivity coefficients show high-frequency sensitivity to $k_{z,\mathrm{sub}}C_{\mathrm{sub}}$, $h_{\mathrm{m}}C_{\mathrm{m}}$, and $G$, while low-frequency sensitivity shifts to $k_{r,\mathrm{sub}}$ and $r_0$. (b1, b2) FDTR phase and normalized amplitude signals in the frequency range of 1 kHz to 50 MHz with a 3.6 μm spot size. Corresponding sensitivity coefficients show low-frequency sensitivity to $k_{r,\mathrm{sub}}/(C_{\mathrm{sub}}r_0^2)$, with increasing high-frequency sensitivity to $k_{z,\mathrm{sub}}$, $h_m$, and $G$. (c1, c2) SPS signals at 5 kHz with a 8 μm spot size, and at 1.2 MHz with a 20 μm spot size. Corresponding sensitivity coefficients show low-frequency sensitivity to $k_{r,\mathrm{sub}}/(C_{\mathrm{sub}}r_0^2)$, and high-frequency sensitivity to $\sqrt{(k_{z,\mathrm{sub}}C_{\mathrm{sub}})/(h_{\mathrm{m}}C_{\mathrm{m}})}$, with varying $G$ sensitivity over time.

Figures 1(a1) and (a2) present the signals measured using TDTR and the corresponding sensitivity coefficients to all parameters in the thermal model. Sensitivity analysis indicates that at a high modulation frequency of 9.8 MHz and a large spot size of 17.4 μm (FIG. 1(a1)), the signals are primarily sensitive to parameters including $k_{z,\mathrm{sub}}C_{\mathrm{sub}}$, $h_{\mathrm{m}}C_{\mathrm{m}}$, and $G$. In contrast, at a low modulation frequency of 0.353 MHz and a small spot size of 3.6 μm (FIG. 1(a2)), the signals become more sensitive to $k_{r,\mathrm{sub}}$ and $r_0$, with reduced sensitivity to $k_{z,\mathrm{sub}}C_{\mathrm{sub}}$, $h_{\mathrm{m}}C_{\mathrm{m}}$, and $G$.

Figures 1(b1) and (b2) show the phase and normalized amplitude signals, respectively, measured using FDTR across a frequency range from 1 kHz to 50 MHz, with a laser spot size of 3.6 μm. The sensitivity coefficients reveal that at low frequencies ($f_0 < 100$ kHz), the signals are predominantly sensitive to the combined parameter $k_{r,\mathrm{sub}}/(C_{\mathrm{sub}}r_0^2)$, as indicated by the sensitivity relationship: $S_{k_{r,\mathrm{sub}}} = -S_{C_{\mathrm{sub}}} = -0.5S_{r_0}$ in this frequency range. At higher frequencies, the sensitivity to parameters like $k_{z,\mathrm{sub}}$, $h_{\mathrm{m}}$, and $G$ increases, and the low-frequency sensitivity relationship between $S_{k_{r,\mathrm{sub}}}$, $S_{C_{\mathrm{sub}}}$, and $S_{r_0}$ no longer applies. Clear relationships between sensitivity coefficients of different parameters are not easily discernible in FDTR experiments.

Figures 1(c1) and (c2) show the measurement signals using the SPS method under two experimental configurations: $f_0 = 5$ kHz, $r_0 = 8$ μm, and $f_0 = 1.2$ MHz, $r_0 = 20$ μm, respectively. Sensitivity analyses reveal that at low frequencies, the signal is primarily sensitive to the combined parameter $k_{r,\mathrm{sub}}/(C_{\mathrm{sub}}r_0^2)$, as indicated by the relationship $S_{k_{r,\mathrm{sub}}} = -S_{C_{\mathrm{sub}}} = -0.5S_{r_0}$. In contrast, at high frequencies, the signal sensitivity shifts to $\sqrt{k_{z,\mathrm{sub}}C_{\mathrm{sub}}}/(h_m C_m)$, where $S_{k_{z,\mathrm{sub}}} = S_{C_{\mathrm{sub}}} = -0.5S_{h_m} = -0.5S_{C_m}$. Additionally, the sensitivity of $G$ initially increases and then decreases with rising normalized time, showing no coupling with other parameters.

Among the three techniques, SPS exhibits the simplest relationships between the sensitivity coefficients of various parameters. This clarity allows us to observe that when the signals are sensitive to $k_{r,\mathrm{sub}}$ and $k_{z,\mathrm{sub}}$, they are essentially sensitive to the combined parameters $k_{r,\mathrm{sub}}/(C_{\mathrm{sub}}r_0^2)$ and $\sqrt{k_{r,\mathrm{sub}}C_{\mathrm{sub}}}/(h_m C_m)$, respectively. In what follows, we continue to explore the full relationships between all parameters and the essential combined parameters in the thermal model. Understanding these intricate relationships is crucial for optimizing parameter fitting in thermoreflectance experiments.

## III. RELATIONSHIPS OF SENSITIVITY COEFFICIENTS IN THERMOREFLECTANCE EXPERIMENTS

### A. Uncovering sensitivity relationships through SVD analysis

We employ singular value decomposition (SVD) to reveal the relationships between sensitivity coefficients of various parameters in thermoreflectance experiments. As an illustrative example, we analyze a hypothetical three-layer system, which includes four parameters for each layer ($k_z, k_r, C, h$), along with interfacial thermal conductances ($G_1$, $G_2$) for the two interfaces and the laser spot size ($r_0$). The specific parameters for this hypothetical system are detailed in TABLE I. To derive broadly applicable conclusions, it is crucial to ensure that the signals are sensitive to all parameters. Verification confirms that the thermal properties listed in Table I satisfy this criterion, particularly with the interface thermal conductance set at $10\ \mathrm{MW/(m^2 \cdot K)}$. The SPS signals across 30 modulation frequencies, equally spaced on a logarithmic scale from 100 Hz to 100 MHz, are analyzed below.

First, we analyze the four parameters of $k_{z1}$, $k_{r1}$, $C_1$, and $h_1$ for layer 1 by examining the sensitivity matrix $\boldsymbol{S}_1 = \left(S_{k_{z1}}, S_{k_{r1}}, S_{C_1}, S_{h_1}\right)$. The sensitivity matrix $\boldsymbol{S}$ is defined as:

$$\boldsymbol{S} = \begin{pmatrix} \vdots \\ \boldsymbol{S}_{f_k} \\ \vdots \end{pmatrix} = \begin{pmatrix} S_{\xi_1} & \cdots & S_{\xi_j} \end{pmatrix} \tag{3}$$

with

TABLE I. Parameters of a hypothetical three-layer system for thermoreflectance experiments, including thermal conductivities ($k_z$, $k_r$), heat capacities ($C$), layer thicknesses ($h$), interfacial thermal conductances ($G_1$, $G_2$), and laser spot size ($r_0$).

| | $k_z$ (W/(m · K)) | $k_r$ (W/(m · K)) | $C$ (MJ/(m$^3$ · K)) | $h$ (nm) | $r_0$ (μm) |
|---|---|---|---|---|---|
| 1(Al) | 150 | 150 | 2.44 | 100 | 8 |
| $G_1$ | | 10 MW/(m$^2$ · K) | | | |
| 2 (Film) | 10 | 100 | 2 | 2000 | |
| $G_2$ | | 10 MW/(m$^2$ · K) | | | |
| 3(Sub) | 100 | 10 | 1.5 | ∞ | |

$$\boldsymbol{S}_{f_k} = \begin{pmatrix} S_{\xi_1,t_1,f_k} & \cdots & S_{\xi_j,t_1,f_k} \\ \vdots & \ddots & \vdots \\ S_{\xi_1,t_i,f_k} & \cdots & S_{\xi_j,t_i,f_k} \end{pmatrix}$$

where $S_{\xi_j,t_i,f_k}$ is the sensitivity coefficient of the SPS signal to the parameter $\xi_j$ at modulation frequency $f_k$ and normalized time $t_i$.

We then decompose $\boldsymbol{S}_1$ using SVD as $\boldsymbol{S}_1 = \boldsymbol{U\Sigma V}^T$. Alternatively, this can be written as:

$$\boldsymbol{S}_1 v_j = u_j \sigma_j \tag{4}$$

where $v_j$ is the *j*-th column of matrix $\boldsymbol{V}$, $u_j$ is the *j*-th column of matrix $\boldsymbol{U}$, and $\sigma_j$ is the *j*-th singular value of $\boldsymbol{S}_1$.

The purpose of applying SVD to $\boldsymbol{S}_1$ is to identify $v_j$ such that $\boldsymbol{S}_1 v_j = 0$. Since $\boldsymbol{U}$ is an orthogonal matrix, the Euclidean norm $\left\|u_j\right\|_2 \equiv 1$, meaning that smaller $\sigma_j$ values indicate that $\boldsymbol{S}v_j$ is closer to the zero vector. When $\max\left(\mathrm{abs}\left(\boldsymbol{S}v_j\right)\right) < 0.001$, we consider $\boldsymbol{S}v_j$ sufficiently close to the zero vector.

Performing SVD on $\boldsymbol{S}_1$ gives the singular value matrix $\boldsymbol{\Sigma} = \mathrm{diag}(33.89, 6.66, 1.23, 0.01)$. Upon review, only $\sigma_4$ is small enough to make $\boldsymbol{S}_1 v_4 \approx 0$, with $v_4 = (-0.51, 0.50, 0.50, -0.50)^T$. Thus, we hypothesize the following relationship:

$$-0.5 S_{k_z} + 0.5 S_{k_r} + 0.5 S_C - 0.5 S_h = 0$$

or equivalently:

$$S_C = S_h + S_{k_z} - S_{k_r} \tag{5}$$

We repeated this analysis for the properties of layers 2 and 3 and found that Eq. (5) holds for all layers.

There is an established principle in sensitivity analysis that, given a signal $R$ as a function of parameters $A$, $B$, and $C$ (i.e., $R = f(A, B, C)$), a sensitivity relationship of the form $S_A = bS_B + cS_C$ serves as a necessary and sufficient condition for expressing $R$ as a function of combined parameters, specifically $R = f(A^b B, A^c C)$. A proof of this principle is provided in **Appendix A**. Applying this principle, Eq. (5) suggests that the signals are primarily sensitive to the combined parameters $k_z C$, $k_r/C$, and $hC$ within each layer.

Additionally, a chain rule applies to the sensitivity coefficients of combined parameters. If the signal $R$ depends on two combined parameters $\alpha$ and $\beta$, where $\alpha = A^a B^b$ and $\beta = A^c B^d$ (with $A$ and $B$ as individual parameters), the chain rule states that the sensitivity coefficients are given by $S_A = aS_\alpha + cS_\beta$, and $S_B = bS_\alpha + dS_\beta$. A proof of this relationship is provided in **Appendix B**. Using this chain rule, and given that the signals are primarily sensitive to the combined parameters $k_z C$, $k_r/C$, and $hC$, we can express the sensitivity coefficients of these combined parameters as:

$$S_{k_z C} = S_{k_z},\ S_{k_r/C} = S_{k_r},\ \text{and}\ S_{hC} = S_h \tag{6}$$

This formulation simplifies sensitivity analysis by directly linking each combined parameter to its respective component, facilitating more straightforward interpretation of the measured signals.

Next, we analyze the sensitivity matrix $\boldsymbol{S}_2$, which incorporates all the parameters involved in this thermal system:

$$\boldsymbol{S}_2 = \left(S_{k_{z1}C_1}, S_{\frac{k_{r1}}{C_1}}, S_{h_1 C_1}, S_{G_1}, S_{k_{z2}C_2}, S_{\frac{k_{r2}}{C_2}}, S_{h_2 C_2}, S_{G_2}, S_{k_{z3}C_3}, S_{\frac{k_{r3}}{C_3}}, S_{r_0}\right).$$

We decompose $\boldsymbol{S}_2$ using SVD as $\boldsymbol{S}_2 = \boldsymbol{U\Sigma V}^T$, resulting in the singular value matrix as:

$$\boldsymbol{\Sigma} = \mathrm{diag}(96.90, 30.36, 23.42, 16.70, 9.52, 3.49, 1.37, 0.28, 0.22, 0.02, 0.01).$$

Upon verification, only the last two singular values, $\sigma_{10}$ and $\sigma_{11}$, are small enough to make $\boldsymbol{S}_2 v_{10} \approx 0$ and $\boldsymbol{S}_2 v_{11} \approx 0$, with:

$$v_{10} = (-0.51, -0.01, -0.25, -0.25, -0.50, 0.01, -0.24, -0.24, -0.50, 0.01, 0.01)^T$$

$$v_{11} = (0, 0.55, 0, 0, 0, 0.55, 0, 0, 0, 0.55, 0.28)^T$$

Converting $[v_{10}, v_{11}]$ into column echelon form yields two simple basis vectors for the null space of $\boldsymbol{S}_2$:

$$(1, 0, 0.5, 0.5, 1, 0, 0.5, 0.5, 1, 0, 0)^T \text{ and } (0, 1, 0, 0, 0, 1, 0, 0, 0, 1, 0.5)^T.$$

From these basis vectors, we propose the following relationships for the sensitivity coefficients of the combined parameters:

$$S_{k_{z1}C_1} + 0.5 S_{h_1 C_1} + 0.5 S_{G_1} + S_{k_{z2}C_2} + 0.5 S_{h_2 C_2} + 0.5 S_{G_2} + S_{k_{z3}C_3} = 0,$$

and

$$S_{\frac{k_{r1}}{C_1}} + S_{\frac{k_{r1}}{C_1}} + S_{\frac{k_{r1}}{C_1}} + 0.5 S_{r_0} = 0$$

Rewritten in a more concise form:

$$2S_{k_{z1}C_1} + S_{h_1 C_1} + S_{G_1} + 2S_{k_{z2}C_2} + S_{h_2 C_2} + S_{G_2} + 2S_{k_{z3}C_3} = 0 \tag{7}$$

$$S_{r_0} = -2S_{\frac{k_{r1}}{C_1}} - 2S_{\frac{k_{r2}}{C_2}} - 2S_{\frac{k_{r3}}{C_3}} = -2\sum_{n=1}^{3} S_{k_{rn}} \tag{8}$$

Combining Eqs. (5), (6), and (7), we derive the following relationship:

$$\sum_{n=1}^{3}\left(S_{k_{zn}} + S_{k_{rn}} + S_{C_n}\right) + \sum_{n=1}^{2} S_{G_n} = 0 \tag{9}$$

Equations (5), (8), and (9) describe the sensitivity relationships for a three-layered system. These relationships can be generalized to a multilayered thermal system with *N* layers. Let the *n*-th layer have an in-plane thermal conductivity $k_{rn}$, out-of-plane thermal conductivity $k_{zn}$, specific heat capacity $C_n$, and thickness $h_n$. $G_n$ denotes the interfacial thermal conductance between the *n*-th and the (*n*+1)-th layer, and $r_0$ is the root-mean-squared (RMS) average of the pump and probe laser spot sizes. The following general relationships apply:

$$S_{h_n} + S_{k_{zn}} = S_{C_n} + S_{k_{rn}} \quad \text{for } n = 1, 2, \ldots, N \tag{10}$$

$$\sum_{n=1}^{N}\left(S_{k_{zn}} + S_{k_{rn}} + S_{C_n}\right) + \sum_{n=1}^{N-1} S_{G_n} = 0 \tag{11}$$

$$S_{r_0} = -2\sum_{n=1}^{N} S_{k_{rn}} \tag{12}$$

TABLE II. Combined parameters and dimensionless forms for each layer in an *N*-layered structure

| Layer | Combined Parameters | Dimensionless Form |
|---|---|---|
| 1 | $\frac{\sqrt{k_{z1}C_1}}{h_1C_1}, \frac{k_{r1}}{C_1r_0^2}$ | $\frac{k_{z1}}{h_1^2C_1f_0}, \frac{k_{r1}}{C_1r_0^2f_0}$ |
| 1/2 | $\frac{G_1}{h_1C_1}$ | $\frac{G_1}{h_1C_1f_0}$ |
| ⋮ | ⋮ | ⋮ |
| *n* | $\frac{\sqrt{k_{zn}C_n}}{h_nC_n}, \frac{\sqrt{k_{zn}C_n}}{h_{n-1}C_{n-1}}, \frac{k_{rn}}{C_nr_0^2}$ | $\frac{k_{zn}}{h_n^2C_nf_0}, \frac{k_{zn}C_n}{h_{n-1}^2C_{n-1}^2f_0}, \frac{k_{rn}}{C_nr_0^2f_0}$ |
| *n*/(*n*+1) | $\frac{G_n}{h_nC_n}$ | $\frac{G_n}{h_nC_nf_0}$ |
| ⋮ | ⋮ | ⋮ |
| *N* | $\frac{\sqrt{k_{zN}C_N}}{h_{N-1}C_{N-1}}, \frac{k_{rN}}{C_Nr_0^2}$ | $\frac{k_{zN}C_N}{h_{N-1}^2C_{N-1}^2f_0}, \frac{k_{rN}}{C_Nr_0^2f_0}$ |

Although the above relationships are derived from a case study of SPS signals, they are also applicable to TDTR and FDTR experiments. For example, in the case of the Al/Sapphire system shown in FIG. 1, Eqs. (10-12) suggest the following relationships among the sensitivity coefficients of various parameters:

$$S_{C_m} = S_{h_m} + S_{k_{z,m}} - S_{k_{r,m}},$$
$$S_{C_{\text{sub}}} = S_{k_{z,\text{sub}}} - S_{k_{r,\text{sub}}},$$
$$S_G = -\left(2S_{k_{z,m}} + S_{h_m} + 2S_{k_{z,\text{sub}}}\right),$$
$$S_{r_0} = -2S_{k_{r,m}} - 2S_{k_{r,\text{sub}}}.$$

These relationships can be easily verified for all the cases shown in FIG. 1.

**B. Derivation of combined parameters**

Combining Eqs. (10), (11), and (6) gives:

$$\sum_{n=1}^{N}\left(2S_{k_{zn}C_n} + S_{h_nC_n}\right) + \sum_{n=1}^{N-1} S_{G_n} = 0 \qquad (13)$$

According to the principle in **Appendix A**, the sensitivity coefficient relationships in Eq. (13) imply that the signals are a function of the following combined parameters:

$$R = f\left(\frac{k_{z1}C_1}{(h_1C_1)^2}, \frac{G_1}{h_1C_1}, \dots, \frac{k_{zn}C_n}{(h_1C_1)^2}, \frac{h_nC_n}{h_1C_1}, \frac{G_n}{h_1C_1}, \dots, \frac{k_{zN}C_N}{(h_1C_1)^2}\right) \qquad (14)$$

To express these parameters in a more intuitive form, Eq. (14) can be rewritten as:

$$R = f\left(\frac{\sqrt{k_{z1}C_1}}{h_1C_1}, \frac{G_1}{h_1C_1}, \dots, \frac{\sqrt{k_{zn}C_n}}{h_nC_n}, \frac{\sqrt{k_{zn}C_n}}{h_{n-1}C_{n-1}}, \frac{G_n}{h_nC_n}, \dots, \frac{\sqrt{k_{zN}C_N}}{h_{N-1}C_{N-1}}\right) \qquad (15)$$

In this expression, the parameter $\frac{\sqrt{k_{zn}C_n}}{h_nC_n}$ represents the ratio of the out-of-plane thermal penetration depth to the thickness of the *n*-th layer, defined as $\frac{d_{p,zn}}{h_n} = \frac{1}{\sqrt{\pi f_0}}\frac{\sqrt{k_{zn}C_n}}{h_nC_n}$. The reciprocal of its square, $\frac{h_n^2C_n}{k_{zn}}$, corresponds to the time required for the heat absorbed by the *n*-th layer to penetrate vertically through this film. Similarly, the reciprocal of the square of $\frac{\sqrt{k_{zn}C_n}}{h_{n-1}C_{n-1}}$, given by $\frac{h_{n-1}^2C_{n-1}^2}{k_{zn}C_n}$, represents the time needed for the heat absorbed by the (*n*-1)-th layer to diffuse through the adjacent *n*-th layer. Additionally, the reciprocal of the parameter $\frac{G_n}{h_nC_n}$ signifies the time required for the heat absorbed by the *n*-th layer to cross the underlying interface. Similar insights have been discussed in prior studies [16,17].

Similarly, combining Eqs. (12) and (6) gives:

$$S_{r_0} = -2\sum_{n=1}^{N} S_{\frac{k_{rn}}{C_n}} \qquad (16)$$

According to the principle in **Appendix A**, the sensitivity coefficient relationships in Eq. (16) suggests that the signals are a function of the following combined parameters:

$$R = f\left(\frac{k_{r1}}{C_1r_0^2}, \dots, \frac{k_{rn}}{C_nr_0^2}, \dots, \frac{k_{rN}}{C_Nr_0^2}\right) \qquad (17)$$

Here, the parameter $\frac{k_{rn}}{C_nr_0^2}$ is derived from the ratio of in-plane thermal penetration depth to the radius of the laser spot, expressed as $\frac{d_{p,rn}}{r_0} = \frac{1}{\sqrt{\pi f_0}}\sqrt{\frac{k_{rn}}{C_nr_0^2}}$. The reciprocal of this parameter represents the time required for the heat absorbed by the *n*-th layer material to diffuse in-plane over an area of $\pi r_0^2$.

By combining Eqs. (15) and (17), TABLE II lists all the combined parameters and their corresponding dimensionless forms for an *N*-layered multilayer structure. These parameters, defined by the heat transfer model, represent the maximum number of independent quantities that can be extracted in the absence of noise. This table is universally applicable to any multilayer sample and any thermoreflectance technique.

Note that some of the combined parameters listed in TABLE II may remain partially coupled and can be further simplified under specific conditions. For instance, if a material layer is extremely thin with a very low out-of-plane thermal conductivity $k_{zn}$, the signal may become insensitive to the layer's specific heat capacity. In such cases, the layer and the adjacent interfaces can be treated as a single equivalent interface, with only the overall cross-plane thermal resistance being measured. Conversely, if the layer has a very high out-of-plane thermal conductivity $k_{zn}$, rendering the signal insensitive to $k_{zn}$, the combined parameters $\frac{\sqrt{k_{zn}C_n}}{h_nC_n}$ and $\frac{\sqrt{k_{zn}C_n}}{h_{n-1}C_{n-1}}$ can be further reduced to $\frac{h_{n-1}C_{n-1}}{h_nC_n}$.

### C. Sensitivity coefficients of combined parameters

We have identified four combined parameters for the $n$-th layer in an $N$-layered system as follows: $\frac{\sqrt{k_{zn}C_n}}{h_nC_n}$, $\frac{\sqrt{k_{zn}C_n}}{h_{n-1}C_{n-1}}$, $\frac{k_{rn}}{C_nr_0^2}$, and $\frac{G_n}{h_nC_n}$, where $2 \leq n \leq N-1$. For $n=1$, the second term involving $h_{n-1}$ is omitted, and for $n=N$, both the first and last terms involving $h_n$ are omitted. Consequently, there are a total of $(4N-3)$ combined parameters for an $N$-layered system. The sensitivity coefficients for these combined parameters can be derived from the sensitivity coefficients of the individual parameters, as detailed below, with proofs provided in **Appendix C**:

$$S_{\frac{\sqrt{k_{zn}C_n}}{h_nC_n}} = \begin{cases} 2S_{k_{z1}};\ n=1 \\ 2S_{k_{z1}} + \sum_{l=2}^{l=n}\left(2S_{k_{zl}} + S_{h_{l-1}} + S_{G_{l-1}}\right);\ 2 \leq n \leq N-1 \end{cases} \quad (18)$$

$$S_{\frac{\sqrt{k_{zn}C_n}}{h_{n-1}C_{n-1}}} = 2S_{k_{zn}} - 2S_{k_{z1}} - \sum_{l=2}^{l=n}\left(2S_{k_{zl}} + S_{h_{l-1}} + S_{G_{l-1}}\right);\ 2 \leq n \leq N \quad (19)$$

$$S_{\frac{G_n}{h_nC_n}} = S_{G_n};\ 1 \leq n \leq N-1 \quad (20)$$

$$S_{\frac{k_{rn}}{C_nr_0^2}} = S_{k_{rn}};\ 1 \leq n \leq N \quad (21)$$

### D. Quantitative assessment of measurable parameters

The combined parameters in Table II represent the theoretical maximum for the thermal response of a multilayered system. However, certain parameters may exhibit low sensitivity or strong inter-parameter coupling, thereby reducing the number of practically measurable parameters, making it essential to assess them for practical applications.

The number of practically measurable parameters can be determined by analyzing the rank of the sensitivity matrix $\boldsymbol{S_c} = \left[S_1 \ldots, S_j, \ldots, S_q\right]_{N_\mathrm{p}\times q}$, where $N_\mathrm{p}$ is the number of signal points and $q$ ($\leq 4N-3$) represents the number of combined parameters under analysis, the sensitivity is defined as in Eq. (1). If the rank of $\boldsymbol{S_c}$ equals $q$, it indicates that the sensitivities of the parameters are linearly independent, and all $q$ parameters can be simultaneously determined by fitting the $N_\mathrm{p}$ signal points. This approach is referred to as the Sensitivity Matrix Reduction Technique (SMART) proposed by Tu and Ong [10].

The rank of $\boldsymbol{S_c}$ can be determined through SVD analysis: $\boldsymbol{S_c} = \boldsymbol{U\Sigma V^T}$, where $\boldsymbol{U}$ and $\boldsymbol{V}$ are orthogonal matrices, and $\boldsymbol{\Sigma}$ is a diagonal matrix of singular values $\sigma_j$: $\boldsymbol{\Sigma} = \mathrm{diag}(\sigma_1 \ldots, \sigma_j, \ldots, \sigma_q)$. The rank of $\boldsymbol{\Sigma}$ is equal to the rank of $\boldsymbol{S_c}$. A threshold of 0.1 is defined: singular values below 0.1 indicate dimensions that contribute negligible information. The number of singular values $\sigma_j > 0.1$ is considered the number of measurable parameters.

The selection of measurable parameters involves identifying the maximal linearly independent vector set from the matrix $\boldsymbol{S_c}$. Since the selection is not unique, a practical approach is as follows: First, select the right singular vectors $v_j$ corresponding to singular values $\sigma_j < 0.1$. Then, identify the rows in these vectors with the highest or second-highest absolute values, as they represent the important parameters in the respective directions. Finally, choose the columns in $\boldsymbol{S_c}$ that correspond to these identified rows. The parameters associated with these columns are treated as known parameters, while the remaining parameters are considered measurable. Once the measurable parameters are determined, the validity of this selection can be further assessed using SVD.

## IV. CASE APPLICATION

To illustrate the practical utility of our framework, we applied the TDTR, FDTR, and SPS techniques to a GaN/Si heterostructure sample, a material system widely used in electronic and optoelectronic applications. This section presents the measurement results and a detailed analysis of the thermal properties extracted using each technique. Through a comparative analysis of TDTR, FDTR, and SPS, we cross-validate the results and showcase the strengths and limitations of each method in extracting thermal conductivity, heat capacity, and interfacial thermal conductance.

### A. Sample description

The GaN/Si heterostructure sample consists of a 1.08 μm-thick GaN layer grown on a silicon substrate via metal-organic chemical vapor deposition (MOCVD). To reduce strain from lattice mismatch, an intermediate layer composed of a 220 nm-thick AlN nucleation layer and a 460 nm-thick AlGaN buffer layer was included between the GaN and Si layers. An 87 nm-thick aluminum (Al) transducer layer was deposited on top of the GaN film to enable thermoreflectance measurements. A high-resolution scanning electron microscope (SEM) cross-sectional image of this GaN/Si heterojunction sample is shown in FIG. 2(a).

A multilayer thermal diffusion model was developed to simulate heat flow through the GaN/Si sample in thermoreflectance experiments. The model includes the GaN layer, AlGaN buffer layer, AlN nucleation layer, and Si substrate, along with the Al transducer, as shown in FIG. 2(b). For each layer, parameters such as thermal conductivity (in-plane and cross-plane), heat capacity, thickness, and interfacial thermal conductance were considered. Given the small thickness of the AlN layer, its in-plane thermal conductivity and heat capacity had minimal impact on the measurements. We thus reduce complexity by treating it as a single interface, where its effective thermal boundary conductance $G_3$ is measured. This approach ensures that the model remains comprehensive without introducing unnecessary complexity. The parameters involved in this heat diffusion model thus include: $k_{r1}$, $k_{z1}$, $C_1$, $h_1$, $G_1$, $k_{r2}$, $k_{z2}$, $C_2$, $h_2$, $G_2$, $k_{r3}$, $k_{z3}$, $C_3$, $h_3$, $G_3$, $k_{r4}$, $k_{z4}$, $C_4$, and $r_0$, totaling 19 parameters. According to the conclusions in TABLE II, only 13 intrinsic combined parameters affect the heat transfer process, as shown in FIG. 2(c).

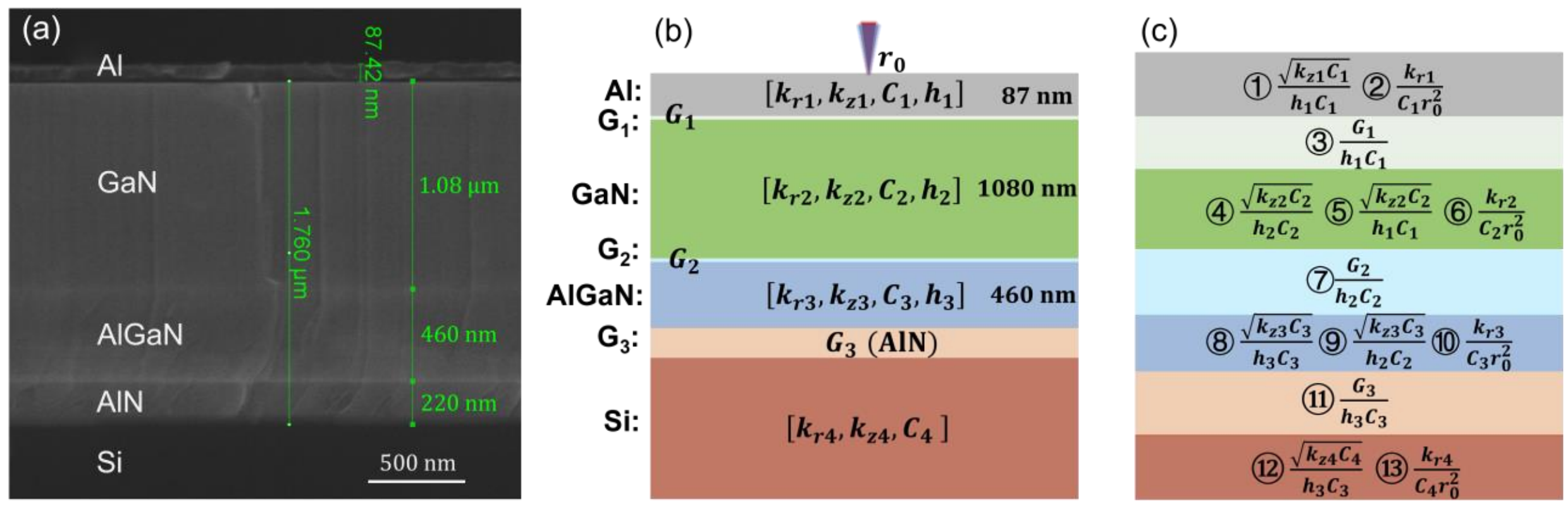

FIG 2. Structural and thermal modeling of a GaN/Si heterostructure. (a) SEM cross-sectional image showing the layered composition from the silicon substrate upwards, including a 220 nm-thick AlN layer, a 460 nm-thick AlGaN layer, a 1.08 µm-thick GaN layer, and an 87 nm-thick aluminum transducer layer. (b) Heat diffusion model of the heterostructure highlighting the primary layers and their respective thermal properties. (c) Depiction of the 13 intrinsic combination parameters that influence the heat transfer process, based on the thermal model established for this specific structure.

### B. TDTR measurement and results

TDTR measurements are performed at three modulation frequencies: 0.611 MHz, 1.084 MHz, and 10.6 MHz. Figure 3(a1-c1) displays the TDTR signals and their best-fit curves at the three modulation frequencies, respectively. Figure 3(a2-c2) presents the sensitivity curves for the 13 combined parameters for each signal group.

Sensitivity plots indicate that all three signal groups are insensitive to parameters ①, ②, ⑩, ⑪, and ⑬. (For clarity, their sensitivity curves are omitted in the plots.) These insensitive parameters correspond to the in-plane and cross-plane thermal conductivities of the Al film ($k_{r1}, k_{z1}$), the in-plane thermal conductivity of the AlGaN layer ($k_{r3}$), the thermal conductance of the AlGaN/Si interface ($G_3$), and the in-plane thermal conductivity of the Si substrate ($k_{r4}$), respectively. Consequently, the following analysis focuses on the 8 remaining parameters: ③, ④, ⑤, ⑥, ⑦, ⑧, ⑨, and ⑫.

We perform SVD on the sensitivity matrix $\boldsymbol{S}_\mathbf{c}$ for these 8 parameters and obtain $\boldsymbol{\Sigma} = \mathrm{diag}(8,3,2,0.6,0.2,0.01,0.0009,0.0003)$. Since $\boldsymbol{\Sigma}$ has 5 singular values greater than 0.1, this indicates that TDTR can reliably measure 5 of these parameters.

Further analysis of the sensitivity plots shows that sensitivity curves for parameters ⑥, ⑦, and ⑧ are highly parallel to that of parameter ⑨, with significantly reduced amplitudes. Therefore, parameters ⑥, ⑦, and ⑧ are excluded from the analysis. Performing SVD on the sensitivity matrix for the remaining parameters (③, ④, ⑤, ⑨, and ⑫) yields $\boldsymbol{\Sigma} = \mathrm{diag}(8.2,2.5,1.6,0.38,0.15)$, with all singular values exceeding 0.1. This indicates that these 5 parameters can be determined from the three groups of TDTR signals.

The above quantitative analysis aligns with a visual assessment of the sensitivity curves. At $f_0 = 10.6$ MHz, the sensitivity curves are most distinct, indicating that the signals are primarily influenced by the combined parameters ③, ④, and ⑤. Each parameter exhibits unique time-dependent sensitivity, enabling their simultaneous determination. Specifically, parameter ③ shows a transition from positive to negative sensitivity at $t_d$ = 0.3 ns, indicating its influence on the signal's slope at this point. Parameter ④ maintains nearly constant sensitivity between $t_d$ = 0.1 ns and $t_d$ = 1 ns, suggesting its primary impact on the signal's magnitude. Meanwhile, parameter ⑤ displays time-dependent sensitivity, implying its role in shaping both the slope and amplitude of the signal. Further analysis of signals at 0.611 MHz and 1.084 MHz reveals distinct relative sensitivities for parameters ⑨ and ⑫, allowing these two parameters to be determined as well.

The five measurable parameters (③, ④, ⑤, ⑨, and ⑫) correspond to $G_1/h_1C_1, \sqrt{k_{z2}C_2}/h_2C_2, \sqrt{k_{z2}C_2}/h_1C_1, \sqrt{k_{z3}C_3}/h_2C_2$, and $\sqrt{k_{z4}C_4}/h_3C_3$, respectively. Therefore, if $h_1, C_1, h_2, C_3, h_3$, and $C_4$ are known, then five parameters including $G_1, k_{z2}, C_2, k_{z3}$, and $k_{z4}$ can be simultaneously determined. Among the required input parameters, the heat capacities $C_1, C_3$, and $C_4$ are taken from the literature, while $h_1$, $h_2$, and $h_3$ are determined via SEM measurements.

Additional parameters, including $k_{r1}, k_{z1}, k_{r2}, G_2, k_{r3}, G_3, k_{z4}$, and $r_0$, are treated as inputs whose exact values have minimal impact on the fitted results. The metal film's in-plane thermal conductivity $k_{r1}$ is calculated from its electrical resistivity, measured using the van der Pauw method and applying the Wiedemann–Franz law. The cross-plane thermal conductivity $k_{z1}$ is assumed equal to $k_{r1}$, and $k_{r2}$ is assumed equal to $k_{z2}$. The laser spot size $r_0$ is measured using the knife-edge method. For the interfacial thermal conductance $G_2$, we assign a reasonable range of $G_2$=$100 \pm 50$ MW/(m$^2 \cdot$ K) based on literature [18]. For $G_3$, we assign a broad range of $200 \pm 100$ MW/(m$^2 \cdot$ K), as we find that when $G_3 > 100$MW/(m$^2 \cdot$ K), the fitting quality and results remain unchanged (see **Supplemental Material Section I** for details). This approach ensures accurate parameter measurement and reliable error analysis for the parameters under investigation.

After sorting out the fitting parameters, we directly fit them using the automatic fitting method (MATLAB function 'lsqnonlin'), and obtain the following values: $G_1 = 120 \pm 5$ MW/(m$^2 \cdot$ K), $k_{z2} = 160 \pm 22$ W/(m $\cdot$ K), $C_2 = 2.7 \pm 0.18$ MJ/(m$^3 \cdot$ K), $k_{z3} = 13.5 \pm 2.5$ W/(m $\cdot$ K), and $k_{z4} = 140 \pm 17$ W/(m $\cdot$ K). The confidence intervals for these results are derived through a comprehensive error analysis, which accounts for both error propagation from input parameters and experimental noise. Detailed uncertainty formulations can be found in references [14,15] and **Supplemental Material Section II**.

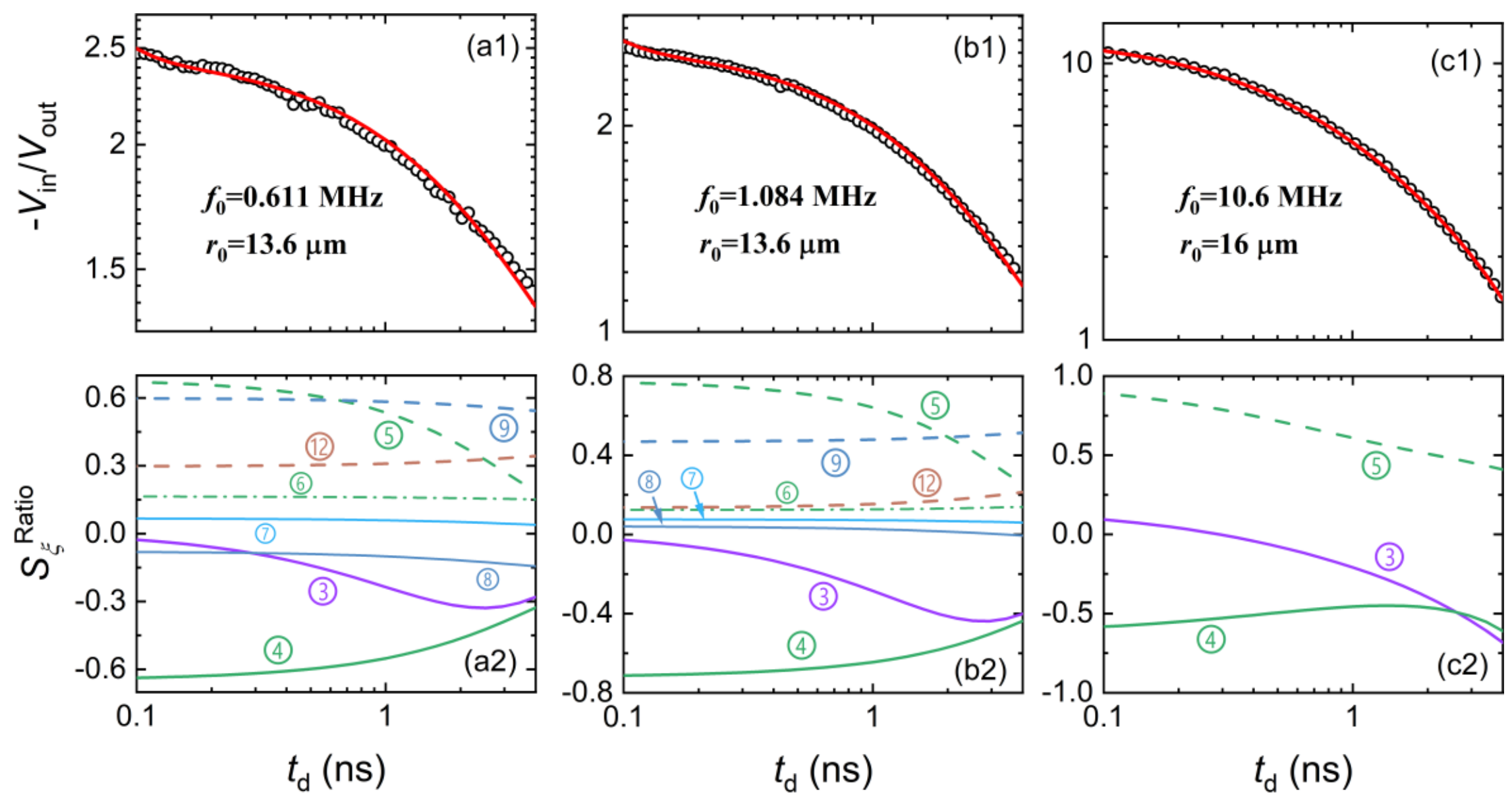

FIG 3. (a1-c1) TDTR signals with corresponding best-fit curves at three modulation frequencies: 0.611 MHz, 1.084 MHz, and 10.6 MHz. (a2-c2) Sensitivity curves for the 13 combined parameters described in FIG. 2(c). To improve clarity, sensitivity curves with magnitudes consistently below 0.05 across all three signal sets are omitted from the plots.

### C. FDTR measurement and results

FDTR measurements are conducted by modulating the pump laser over a frequency range of 10 kHz to 50 MHz. Thermal properties are extracted by analyzing the phase and amplitude of the thermoreflectance signals. Figure 4 shows the measured signals and their corresponding sensitivity curves. Specifically, Figures 4(a1) and 4(b1) display the same FDTR phase signals, along with their best-fit results, presented on linear and logarithmic scales, respectively, while Figures. 4(a2) and 4(b2) present the sensitivity curves, with different sensitivity definitions, for the 13 combined parameters in the system. Figures 4(c1) and 4(c2) present the FDTR normalized amplitude signals and the corresponding sensitivity curves, respectively.

In our data processing, we primarily fit the phase signals in the frequency range of 10 kHz to 30 MHz, as this range offers higher sensitivity and reliability for parameter extraction. Signals above 30 MHz, which tend to exhibit larger phase shifts and reduced accuracy, are excluded from the fitting. Amplitude signals, due to their strong parameter coupling, are used to validate the phase fitting results. This approach helps minimize errors caused by high-frequency discrepancies and ensures robust determination of thermal properties.

Based on the sensitivity curves, the signals are primarily sensitive to parameters ③, ④, ⑤, ⑥, ⑧, ⑨, ⑫, and ⑬. We perform SVD on the sensitivity matrix $\boldsymbol{S}_{\mathbf{c}}$ for these 8 combined parameters and obtain $\boldsymbol{\Sigma} = \mathrm{diag}(8,4,1,0.8,0.2,0.08,0.03,0.008)$, indicating that FDTR can reliably measure 5 of these parameters. By excluding parameters with high coupling coefficients, we select parameters ③, ⑤, ⑥, ⑨, and ⑬ for further SVD analysis, resulting in $\boldsymbol{\Sigma} = \mathrm{diag}(5.8,2.6,1.3,0.64,0.15)$, which confirms that these parameters have been successfully decoupled.

The 5 measurable parameters (③, ⑤, ⑥, ⑨, and ⑬) involve $G_1/h_1C_1$, $\sqrt{k_{z2}C_2}/h_1C_1$, $k_{r2}/C_2r_0^2$, $\sqrt{k_{z3}C_3}/h_2C_2$, and $k_{r4}/C_4r_0^2$. Meanwhile, the other 3 highly sensitive but coupled parameters (④, ⑧, and ⑫) are $\sqrt{k_{z2}C_2}/h_2C_2$, $\sqrt{k_{z3}C_3}/h_3C_3$, and $\sqrt{k_{z4}C_4}/h_3C_3$. To decouple these parameters, we combine them with measurable ones. Specifically, combining the coupled parameter ④ with the measurable parameter ⑤ yields the ratio $h_1C_1/h_2C_2$, and combining ⑧with ⑨ gives the ratio $h_2C_2/h_3C_3$. The coupled parameter ⑫ involving $k_{z4}$ can be simplified by assuming isotropic behavior in the Si substrate, allowing us to set $k_{z4} = k_{r4} = k_4$, thereby reducing the complexity of the problem. Therefore, with eight known parameters—$h_1$, $C_1$, $h_2$, $C_2$, $h_3$, $C_3$, $C_4$, and $r_0$—we can accurately determine the five parameters: $G_1$, $k_{z2}$, $k_{r2}$, $k_{z3}$, and $k_4$. Among the required inputs, the thicknesses $h_1$, $h_2$, and $h_3$ can be measured using SEM, the specific heat capacities $C_1$, $C_2$, $C_3$, and $C_4$ can be obtained from standard databases, and the spot size $r_0$ can be measured using the knife-edge method.

The remaining five parameters—$k_{z1}$, $k_{r1}$, $G_2$, $k_{r3}$, and $G_3$—have little impact on the fitting results and thus are handled using the same simplified approach as in TDTR experiments.

We use an automatic fitting method (MATLAB's lsqnonlin function) to determine the five parameters: $G_1$, $k_{z2}$, $k_{r2}$, $k_{z3}$, and $k_4$. The fitted results are: $G_1 = 129 \pm 7\ \mathrm{MW/(m^2 \cdot K)}$, $k_{z2} = 136 \pm 14\ \mathrm{W/(m \cdot K)}$, $k_{r2} = 153 \pm 12\ \mathrm{W/(m \cdot K)}$, $k_{z3} = 12.2 \pm 1.8\ \mathrm{W/(m \cdot K)}$, and $k_4 = 139 \pm 5\ \mathrm{W/(m \cdot K)}$.

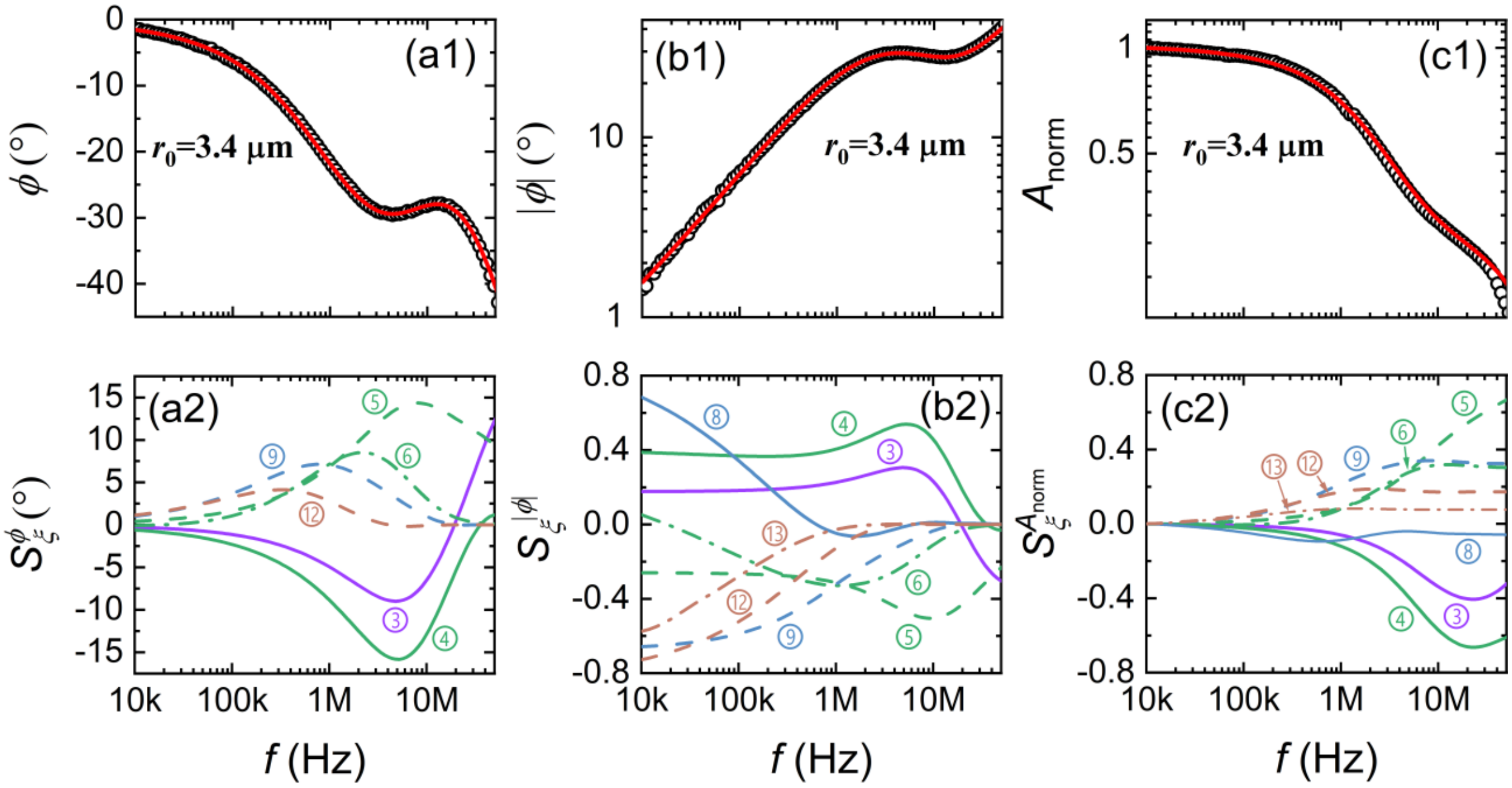

FIG 4. FDTR signal analysis and sensitivity to parameters. (a1-c1) Measured phase and amplitude signals from FDTR with corresponding best-fit results. The $y$-axis in (a1) is linear, while the $y$-axes in (b1) and (c1) are logarithmic. (a2-c2) Sensitivity curves for 13 combined parameters. In (a2), sensitivity curves with magnitudes below 1° are excluded, while in (b2) and (c2), sensitivity curves with magnitudes below 0.05 are omitted to enhance clarity.

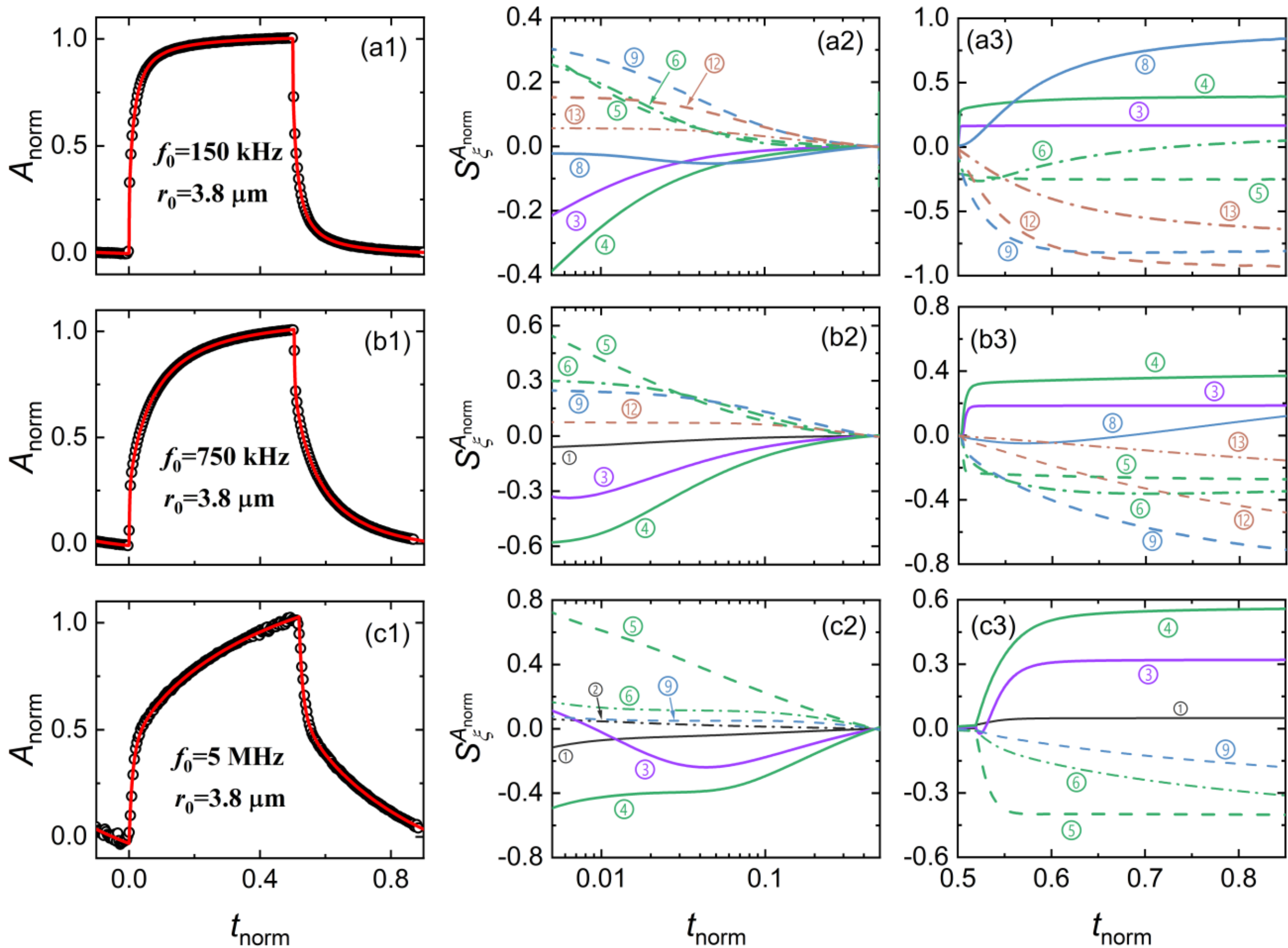

FIG 5. SPS measurement signals and sensitivity analysis. Panels (a1-c1) show the measured SPS signals at 150 kHz, 750 kHz, and 5 MHz, respectively, while (a2-c2) display the sensitivity curves of the combined parameters during the heating phase, and (a3-c3) during the cooling phase. Sensitivity curves with magnitudes consistently below 0.05 for all three signal sets are omitted from the plots to enhance clarity.

### D. SPS measurement and results

We further apply the recently developed SPS technique to measure this sample using three different frequencies: 5 MHz, 750 kHz, and 150 kHz. The high stability of the amplitude signal and the richness of the sensitivity allow SPS to simultaneously extract more parameters. Figure 5 presents the SPS measurement signals in (a1-c1), accompanied by the sensitivity curves of the combined parameters during the heating phase (a2-c2) and the cooling phase (a3-c3).

Based on the sensitivity curves, the signals exhibit high sensitivity to parameters ③, ④, ⑤, ⑥, ⑧, ⑨, ⑫, and ⑬. We apply SVD to the sensitivity matrix $\boldsymbol{S}_\mathbf{c}$ for these parameters and obtain $\boldsymbol{\Sigma} = \mathrm{diag}(32,16,6,2,0.7,0.4,0.2,0.06)$. Since $\boldsymbol{\Sigma}$ has 7 singular values greater than 0.1, this indicates that SPS can reliably measure up to 7 of these parameters. Upon closer inspection, we find that parameter ⑨ is highly coupled with the others. After excluding it from the analysis, we focus on the remaining parameters (③, ④, ⑤, ⑥, ⑧, ⑫, and ⑬) and perform further SVD analysis, yielding $\boldsymbol{\Sigma} = \mathrm{diag}(27,16,5,2,0.6,0.4,0.2)$, confirming successfully decoupled of these parameters.

This quantitative analysis is supported by a visual examination of the sensitivity curves. Specifically, Figure 5(c2) shows that the 5 MHz signals are primarily sensitive to the combined parameters ③, ④, and ⑤, with their sensitivity curves displaying distinct trends over time. In Figure 5(a3), only the sensitivity of parameter ⑥ decreases gradually over time. Moreover, in Figures 5(a3) and 5(b3), the sensitivity of parameter ⑧ transitions from negative to positive, while the sensitivity ratio between parameters ⑫ and ⑬ varies with frequency. As a result, all these 7 combined parameters can be simultaneously determined.

These seven measurable combined parameters correspond to 7 fitting parameters ($G_1$, $k_{z2}$, $C_2$, $k_{r2}$, $k_{z3}$, $k_4$, and $C_4$) and 6 input parameters ($h_1$, $C_1$, $h_2$, $h_3$, $C_3$, and $r_0$). The isotropic nature of the Si substrate allows us to treat $k_{z4} = k_{r4} = k_4$, enabling the simultaneous determination of $k_4$ and $C_4$ from the fitted results of $k_{z4}C_4$ and $k_{r4}/C_4$. The remaining five parameters— $k_{z1}$, $k_{r1}$, $G_2$, $k_{r3}$, and $G_3$—have minimal impact on the fitting results and are thus handled using the same simplified approach employed in TDTR and FDTR experiments.

We use an automatic fitting program (MATLAB function 'lsqnonlin') to best fit the three sets of signals, with $G_1$, $k_{z2}$, $C_2$, $k_{r2}$, $k_{z3}$, $k_4$, and $C_4$ treated as fitting parameters. The fitting results are as follows: $G_1 = 125 \pm 7\ \mathrm{MW/(m^2 \cdot K)}$, $k_{z2} = 145 \pm 13\ \mathrm{W/(m \cdot K)}$, $C_2 = 2.74 \pm 0.19\ \mathrm{MJ/(m^3 \cdot K)}$, $k_{r2} = 146 \pm 15\ \mathrm{W/(m \cdot K)}$, $k_{z3} = 12.3 \pm 3.1\ \mathrm{W/(m \cdot K)}$, $k_4 = 143 \pm 7\ \mathrm{W/(m \cdot K)}$, and $C_4 = 1.67 \pm 0.26\ \mathrm{MJ/(m^3 \cdot K)}$.

### E. Summary and comparison

TABLE III provides a comparative summary of the thermal properties extracted using TDTR, FDTR, and SPS. Unlike conventional practices that only fit $G_1$ and $k_{z2}$, these thermoreflectance techniques demonstrated enhanced capabilities. Specifically, TDTR and FDTR can successfully extract five parameters simultaneously, while SPS can extract up to seven parameters.

The thermal properties obtained using the three techniques show strong consistency. For example, all three methods reveal that the thermal conductance $G_1$ for the Al/GaN interface ranges from 120 to 129 $\mathrm{MW/(m^2 \cdot K)}$. The measured in-plane and cross-plane thermal conductivities ($k_{r2}$ and $k_{z2}$) of the GaN film are similar, indicating weak anisotropy, which can be attributed to two counteracting factors: vertical dislocations that primarily reduce in-plane thermal conductivity and boundary scattering due to film thickness that predominantly reduces cross-plane thermal conductivity. The thermal conductivity of AlGaN obtained from all three methods is consistent, albeit with relatively high uncertainties of 15% to 25%. This variability arises primarily from the 50% error margin set for $G_2$, which is strongly coupled with $k_{z3}$. While the thermal conductivities of the GaN film ($k_{r2}$ and $k_{z2}$), AlGaN film ($k_{z3}$), and Al/GaN interface thermal conductance ($G_1$) are greatly influenced by sample quality and preparation conditions, making direct comparisons with literature data challenging, the heat capacities of the GaN film ($C_2$) and Si substrate ($C_4$), as well as the thermal conductivity of the Si substrate ($k_4$), are relatively stable and serve as reliable benchmarks for verifying the accuracy of our results. The thermal conductivity of single-crystal silicon reported in the literature is $142 \pm 3\ \mathrm{W/(m \cdot K)}$ [14,19], and the specific heat capacity of GaN is $2.63 \pm 0.08\ \mathrm{MJ/(m^3 \cdot K)}$ [20], both of which are highly consistent with the results of this study.

In the literature, Mitterhuber *et al.* [18] also employed TDTR (PicoTR, Netzsch) to measure the thermal conductivity of GaN/Si heterostructures. To assess the thermal conductivities of various thin-film layers, they prepared four samples by grinding and polishing to expose specific layers: the AlN nucleation layer, the $Al_{0.32}Ga_{0.68}N$ transition layer, the GaN buffer layer, and the $Al_{0.17}Ga_{0.83}N$ top barrier layer. Each sample surface was coated with a ~100 nm thick Pt film as a transducer layer. Their measurements yielded a thermal conductivity of $220 \pm 38\ \mathrm{W/(m \cdot K)}$ for the 1.07 μm thick GaN layer, $11.2 \pm 0.7\ \mathrm{W/(m \cdot K)}$ for the 423 nm thick $Al_{0.32}Ga_{0.68}N$ layer, and $9.7 \pm 0.6\ \mathrm{W/(m \cdot K)}$ for the 65 nm thick $Al_{0.17}Ga_{0.83}N$ layer. These results show close agreement with the findings of this study.

TABLE III. Comparative analysis of measurement results from TDTR, FDTR, and SPS.

| | Measured thermal properties | TDTR | FDTR | SPS |
|---|---|---|---|---|
| Al/GaN | $G_1$ ($\mathrm{MW/(m^2 \cdot K)}$) | $120 \pm 5$ | $129 \pm 7$ | $125 \pm 7$ |
| GaN | $k_{z2}$ ($\mathrm{W/(m \cdot K)}$) | $160 \pm 22$ | $136 \pm 14$ | $145 \pm 13$ |
| | $k_{r2}$ ($\mathrm{W/(m \cdot K)}$) | - | $153 \pm 12$ | $146 \pm 15$ |
| | $C_2$ ($\mathrm{MJ/(m^3 \cdot K)}$) | $2.7 \pm 0.18$ | - | $2.74 \pm 0.19$ |
| AlGaN | $k_3$ ($\mathrm{W/(m \cdot K)}$) | $13.5 \pm 2.5$ | $12.2 \pm 1.8$ | $12.3 \pm 3.1$ |
| Si | $k_4$ ($\mathrm{W/(m \cdot K)}$) | $140 \pm 17$ | $139 \pm 5$ | $143 \pm 7$ |
| | $C_4$ ($\mathrm{MJ/(m^3 \cdot K)}$) | - | - | $1.67 \pm 0.26$ |

## V. CONCLUSIONS

In this study, we developed a systematic framework using SVD to decouple complex parameter interdependencies in thermoreflectance experiments. Key combined parameters were identified, and essential dimensionless numbers governing thermoreflectance signals were established. By applying the framework to TDTR, FDTR, and SPS techniques, we demonstrated its ability to improve the precision of thermal property measurements in multilayer systems. Measurements conducted on a GaN/Si heterostructure using these three techniques validated this framework, showing strong consistency across the results.

The proposed framework significantly enhances the accuracy and reliability of thermoreflectance-based thermal characterization, providing a robust foundation for improved measurement practices in material science, where precise thermal measurements are essential for optimizing material performance and ensuring the reliability of next-generation devices.

## ACKNOWLEDGMENTS

The authors would like to express their sincere gratitude to Professor David Cahill from UIUC and the anonymous reviewers for their valuable and constructive feedback on this work. They also wish to thank Professor Junjun Wei of Beijing University of Science and Technology for providing the GaN film sample used in this study, Dr. Weidong Zheng from the Shandong Institute of Advanced Technology for performing the TDTR measurements, and Dr. Jun Su from the Center of Optoelectronic Micro & Nano Fabrication and Characterization Facility at the Wuhan National Laboratory for Optoelectronics, Huazhong University of Science and Technology, for his assistance with the SEM testing. This research was supported by the National Natural Science Foundation of China (NSFC) under Grant No. 52376058.

## APPENDIX A: GIVEN $R = F(A, B, C)$. PROVE THAT $S_A = bS_B + cS_C$ IS THE NECESSARY AND SUFFICIENT CONDITION FOR $R = F(A^bB, A^cC)$.

First, prove the necessity. Assuming $R = F(A^bB, A^cC)$, the sensitivity coefficients for $A$, $B$, and $C$ are:

$$S_A = \frac{A}{R}\frac{\partial R}{\partial A} = \frac{A}{R}\left(bA^{b-1}BF_1'(A^bB, A^cC) + cA^{c-1}CF_2'(A^bB, A^cC)\right) \quad \text{(A1)}$$

$$S_B = \frac{B}{R}\frac{\partial R}{\partial B} = \frac{B}{R}A^bF_1'(A^bB, A^cC) \quad \text{(A2)}$$

$$S_C = \frac{C}{R}\frac{\partial R}{\partial C} = \frac{C}{R}A^cF_2'(A^bB, A^cC) \quad \text{(A3)}$$

From Eqs. (A1-A3), we have:

$$bS_B + cS_C = b\frac{B}{R}A^bF_1'(A^bB, A^cC) + c\frac{C}{R}A^cF_2'(A^bB, A^cC) = S_A \quad \text{(A4)}$$

Then prove sufficiency. Assuming $S_A = bS_B + cS_C$ and supposing that $R$ cannot be solely represented by two parameters, consider a third parameter $D$, let $D = A$. Thus, $R = F(A^bB, A^cC, D)$, and the sensitivity coefficients for $A$, $B$, and $C$ are:

$$S_A = \frac{A}{R}\frac{\partial R}{\partial A} = \frac{A}{R}\left(bA^{b-1}BF_1'(A^bB, A^cC, D) + cA^{c-1}CF_2'(A^bB, A^cC, D) + F_3'(A^bB, A^cC, D)\right) \quad \text{(A5)}$$

$$S_B = \frac{B}{R}\frac{\partial R}{\partial B} = \frac{B}{R}A^bF_1'(A^bB, A^cC, D) \quad \text{(A6)}$$

$$S_C = \frac{C}{R}\frac{\partial R}{\partial C} = \frac{C}{R}A^cF_2'(A^bB, A^cC, D) \quad \text{(A7)}$$

Substituting Eqs. (A5-A7) into $S_A = bS_B + cS_C$, we obtain:

$$F_3'(A^bB, A^cC, D) = 0 \quad \text{(A8)}$$

Due to the arbitrariness of $A$, $B$, and $C$, $R$ does not change with the third parameter, disproving the assumption, thus $R$ is a function of only $A^bB$ and $A^cC$, i.e., $R = F(A^bB, A^cC)$.

## APPENDIX B: THE CHAIN RULE FOR SENSITIVITY COEFFICIENTS

Suppose the signal $R$ is given by $R = F(\alpha, \beta)$, where $\alpha$ and $\beta$ are two combined parameters determining $R$. Let $\alpha = A^aB^b$ and $\beta = A^cB^d$, where $A, B$ are individual parameters. According to the chain rule:

$$\frac{\partial R}{\partial A} = \frac{\partial \alpha}{\partial A}\frac{\partial R}{\partial \alpha} + \frac{\partial \beta}{\partial A}\frac{\partial R}{\partial \beta} = aA^{a-1}B^b\frac{\partial R}{\partial \alpha} + cA^{c-1}B^d\frac{\partial R}{\partial \beta} \quad \text{(B1)}$$

Based on the definition of sensitivity, we have:

$$S_A = \frac{A}{R}\frac{\partial R}{\partial A} = aA^{a-1}B^b\frac{A}{R}\frac{\partial R}{\partial \alpha} + cA^{c-1}B^d\frac{A}{R}\frac{\partial R}{\partial \beta} = a\frac{\alpha}{R}\frac{\partial R}{\partial \alpha} + c\frac{\beta}{R}\frac{\partial R}{\partial \beta} = aS_\alpha + cS_\beta \quad \text{(B2)}$$

Equation (B2) represents the chain rule for sensitivity coefficients. To calculate the sensitivity of an individual parameter, simply add the power indices of that parameter in each combined parameter as the coefficients of the sensitivity coefficients of the combined parameters.

## APPENDIX C: THE FORMULAS FOR THE SENSITIVITY COEFFICIENTS OF COMBINED PARAMETERS

According to the chain rule and the definition of sensitivity, we have:

$$S_{k_{z1}} = \frac{1}{2}S_{\frac{\sqrt{k_{z1}C_1}}{h_1C_1}} \quad \text{(C1)}$$

$$S_{G_n} = S_{\frac{G_n}{h_nC_n}}; 1 \le n \le N-1 \quad \text{(C2)}$$

$$S_{k_{zn}} = \frac{1}{2}S_{\frac{\sqrt{k_{zn}C_n}}{h_nC_n}} + \frac{1}{2}S_{\frac{\sqrt{k_{zn}C_n}}{h_{n-1}C_{n-1}}}; 2 \le n \le N-1 \quad \text{(C3)}$$

$$S_{h_n} = -S_{\frac{\sqrt{k_{zn}C_n}}{h_nC_n}} - S_{\frac{\sqrt{k_{z(n+1)}C_{(n+1)}}}{h_nC_n}} - S_{\frac{G_n}{h_nC_n}}; 1 \le n \le N-1 \quad \text{(C4)}$$

$$S_{k_{zN}} = \frac{1}{2}S_{\frac{\sqrt{k_{zN}C_N}}{h_{N-1}C_{N-1}}} \quad \text{(C5)}$$

$$S_{k_{rn}} = S_{\frac{k_{rn}}{C_nr_0^2}}; 1 \le n \le N-1 \quad \text{(C6)}$$

From Eq.(C1):

$$S_{\frac{\sqrt{k_{z1}C_1}}{h_1C_1}} = 2S_{k_{z1}} \quad \text{(C7)}$$

From Eq.(C2):

$$S_{\frac{G_n}{h_nC_n}} = S_{G_n}; 1 \le n \le N-1 \quad \text{(C8)}$$

From Eq.(C5):

$$S_{\frac{\sqrt{k_{zN}C_N}}{h_{N-1}C_{N-1}}} = 2S_{k_{zN}} \quad \text{(C9)}$$

From Eq.(C6):

$$S_{\frac{k_{rn}}{C_nr_0^2}} = S_{k_{rn}}; 1 \le n \le N \quad \text{(C10)}$$

Adjusting Eqs. (C2, C4) with $n$ changed to $n-1$, we find:

$$S_{h_{n-1}} = -S_{\frac{\sqrt{k_{zn-1}C_{n-1}}}{h_{n-1}C_{n-1}}} - S_{\frac{\sqrt{k_{z(n)}C_{(n)}}}{h_{n-1}C_{n-1}}} - S_{G_{n-1}}; 2 \le n \le N \quad \text{(C11)}$$

Combining Eqs.(C3, C11) yields the recursive formula for $S_{\frac{\sqrt{k_{zn}C_n}}{h_nC_n}}$:

$$S_{\frac{\sqrt{k_{zn}C_n}}{h_nC_n}} = S_{\frac{\sqrt{k_{zn-1}C_{n-1}}}{h_{n-1}C_{n-1}}} + 2S_{k_{zn}} + S_{h_{n-1}} + S_{G_{n-1}}; 2 \le n \le N-1 \quad \text{(C12)}$$

From Eqs. (C7, C12):

$$S_{\frac{\sqrt{k_{zn}C_n}}{h_nC_n}} = \begin{cases} 2S_{k_{z1}};\ n = 1 \\ 2S_{k_{z1}} + \sum_{l=2}^{l=n}\left(2S_{k_{zl}} + S_{h_{l-1}} + S_{G_{l-1}}\right);\ 2 \le n \le N-1 \end{cases} \quad \text{(C13)}$$

Combining Eqs. (C3, C9, C13):

$$S_{\frac{\sqrt{k_{zn}C_n}}{h_{n-1}C_{n-1}}} = 2S_{k_{zn}} - 2S_{k_{z1}} - \sum_{l=2}^{l=n}\left(2S_{k_{zl}} + S_{h_{l-1}} + S_{G_{l-1}}\right);\ 2 \le n \le N \quad \text{(C14)}$$

Equation (C8) is the formula for calculating the sensitivity of the combined parameter $\frac{G_n}{h_nC_n}$, Eq. (C10) is for the combined parameter $\frac{k_{rn}}{C_nr_0^2}$, Eq. (C13) is for the combined parameter $\frac{\sqrt{k_{zn}C_n}}{h_nC_n}$, and Eq. (C14) is for the combined parameter $\frac{\sqrt{k_{zn}C_n}}{h_{n-1}C_{n-1}}$.

[1] C. A. Gadre *et al.*, Nanoscale imaging of phonon dynamics by electron microscopy, Nature **606**, 292 (2022).

[2] P.-A. Krochin-Yepez, U. Scholz, and A. Zimmermann, in *Photonics* (MDPI, 2020), p. 6.

[3] M. Elhajhasan *et al.*, Optical and thermal characterization of a group-III nitride semiconductor membrane by microphotoluminescence spectroscopy and Raman thermometry, Phys. Rev. B **108**, 235313 (2023).

[4] D. G. Cahill, Analysis of heat flow in layered structures for time-domain thermoreflectance, Rev. Sci. Instrum. **75**, 5119 (2004).

[5] P. Jiang, X. Qian, and R. Yang, Time-domain thermoreflectance (TDTR) measurements of anisotropic thermal conductivity using a variable spot size approach, Rev. Sci. Instrum. **88**, 074901 (2017).

[6] P. Jiang, X. Qian, and R. Yang, A new elliptical-beam method based on time-domain thermoreflectance (TDTR) to measure the in-plane anisotropic thermal conductivity and its comparison with the beam-offset method, Rev. Sci. Instrum. **89**, 094902 (2018).

[7] A. J. Schmidt, R. Cheaito, and M. Chiesa, A frequency-domain thermoreflectance method for the characterization of thermal properties, Rev. Sci. Instrum. **80**, 094901, 094901 (2009).

[8] J. Liu, J. Zhu, M. Tian, X. Gu, A. Schmidt, and R. Yang, Simultaneous measurement of thermal conductivity and heat capacity of bulk and thin film materials using frequency-dependent transient thermoreflectance method, Rev. Sci. Instrum. **84**, 034902, 034902 (2013).

[9] D. Rodin and S. K. Yee, Simultaneous measurement of in-plane and through-plane thermal conductivity using beam-offset frequency domain thermoreflectance, Rev. Sci. Instrum. **88**, 014902 (2017).

[10] J. Tu and W.-L. Ong, A universal sensitivity matrix reduction technique (SMART) to uncover governing thermal transport relationships, Int. J. Heat Mass Transf. **206**, 123949 (2023).

[11] P. Jiang, X. Qian, and R. Yang, Tutorial: Time-domain thermoreflectance (TDTR) for thermal property characterization of bulk and thin film materials, J Appl Phys **124**, 161103 (2018).

[12] C. Wei, X. Zheng, D. G. Cahill, and J. C. Zhao, Invited article: micron resolution spatially resolved measurement of heat capacity using dual-frequency time-domain thermoreflectance, Rev Sci Instrum **84**, 071301 (2013).

[13] E. Ziade, Wide bandwidth frequency-domain thermoreflectance: Volumetric heat capacity, anisotropic thermal conductivity, and thickness measurements, Rev. Sci. Instrum. **91**, 124901 (2020).

[14] T. Chen, S. Song, Y. Shen, K. Zhang, and P. Jiang, Simultaneous measurement of thermal conductivity and heat capacity across diverse materials using the square-pulsed source (SPS) technique, Int. Commun. Heat Mass Transf. **158**, 107849 (2024).

[15] T. Chen, S. Song, R. Hu, and P. Jiang, Comprehensive measurement of three-dimensional thermal conductivity tensor using a beam-offset square-pulsed source (BO-SPS) approach, Int J Therm Sci **207**, 109347 (2025).

[16] O. M. Wilson, X. Hu, D. G. Cahill, and P. V. Braun, Colloidal metal particles as probes of nanoscale thermal transport in fluids, Phys. Rev. B **66**, 224301 (2002).

[17] R. Wilson, J. P. Feser, G. T. Hohensee, and D. G. Cahill, Two-channel model for nonequilibrium thermal transport in pump-probe experiments, Phys. Rev. B Condens. Matter Mater. Phys. **88**, 144305 (2013).

[18] L. Mitterhuber, R. Hammer, T. Dengg, and J. Spitaler, Thermal characterization and modelling of AlGaN-GaN multilayer structures for HEMT applications, Energies **13**, 2363 (2020).

[19] A. V. Inyushkin, A. N. Taldenkov, A. M. Gibin, A. V. Gusev, and H. J. Pohl, On the isotope effect in

thermal conductivity of silicon, physica status solidi (c) **1**, 2995 (2004).
[20] B. A. Danilchenko, T. Paszkiewicz, S. Wolski, A. Jeżowski, and T. Plackowski, Heat capacity and phonon mean free path of wurtzite GaN, Appl. Phys. Lett. **89**, 061901 (2006).